\documentstyle[12pt,epsf]{article}
\begin{document}
\title{Investigating liquid surfaces down to the nanometer scale 
using grazing incidence x-ray scattering.}
\author{C. Fradin$^1$, 
A. Braslau$^1$, D. Luzet$^1$, M. Alba$^1$
C. Gourier$^{1,*}$, 
\underline{J. Daillant}$^1$, \\
G. Gr\"ubel$^2$, G. Vignaud$^2$, J.F. Legrand$^{2,3}$,
J. Lal$^2$,\\ J.M. Petit$^3$, F. Rieutord$^3$. \\
$^1$ Service de Physique de l'Etat Condens\'{e}, Orme des Merisiers,
CEA Saclay, \\
F-91191 Gif-sur-Yvette Cedex, France.\\
$^2$ European Synchrotron Radiation Facility, B.P. 220, \\
F-38043 Grenoble
Cedex, France.\\
$^3$ CEA/DRFMC/SI3M, F-38054 Grenoble Cedex 9, France.}
\maketitle

\noindent{\it Author for correspondence:} J. Daillant, 
\\
Service de Physique de l'Etat Condens\'{e},\\
Orme des Merisiers, CEA Saclay,\\
F-91191 Gif-sur-Yvette Cedex, France.\\
\noindent fax: 33 1 69 08 81 57.\\
\noindent e-mail: daillant@spec.saclay.cea.fr\\

\noindent {\it Fifth international conference on surface x-ray and neutron
scattering.}

\vskip 2truecm
\begin{abstract}
Grazing incidence x-ray surface scattering has been used to investigate
liquid surfaces down to the molecular scale. The free surface of
water is well described by the capillary wave model ($<z(q)z(-q)> 
\propto q^{-2}$ spectrum) up to wave-vectors $> 10^8 m^{-1}$. 
At larger wave-vectors near surface acoustic waves must be taken
into account. \\
When the interface is bounded by a surfactant monolayer, it exhibits 
a bending stiffness and the bending rigidity modulus can be measured. 
However, bending effects generally cannot be described using
the Helfrich Hamiltonian and the characteristic exponent in the roughness 
power spectrum can be smaller than 4.
Finally, upon compression, tethered monolayers formed on a subphase containing 
divalent ions are shown to buckle in the third dimension with a characteristic 
wavelength on the order of $ 10^8 m^{-1}$. 
\end{abstract}

\vskip 0.5truecm
\noindent{\it Keywords: } Liquid surfaces, Surfactant monolayers.\\
\newpage

\section{Introduction.}
The structure and dynamics of liquid surfaces and interfaces has attracted
much attention since the work of van der Waals \cite{Widom}.
The distinctive property of liquid surfaces is a logarithmic  
divergence with distance of the roughness, only limited by gravity 
or the finite size of the surface. This behavior is well described 
by the capillary wave model of Buff, Lovett and Stillinger \cite{Buff}
which yields a height-height correlation function:
$<z({\bf 0}) z({\bf r})> = {k_BT / (2 \pi \gamma)}\> K_0\left(r \sqrt{\Delta
\rho g / \gamma} \right)$, where ${\bf r}$ is the vector joining two 
points in the plane of the surface, $\gamma$ is the surface tension, 
and $\Delta \rho$ the difference in density between the two phases 
separated by the interface. $K_0$ is the modified Bessel function of 
second kind of order 0: $lim K_0(x)_{x \to 0} \approx -ln x$. 
This height-height correlation function is the Fourier transform of the 
following spectrum:
\begin{equation}
<z({\bf q}) z({\bf -q})> = {1 \over A} {k_B T \over \Delta \rho g + \gamma q^2}
\label{spectrum}
\end{equation}
where $A$ is the interface area. 
This spectrum has been well characterized for many liquid surfaces 
by light scattering down to 
wave-lengths in the micrometer range \cite{Loudon} and is valid is the limit
of small in-plane momentum ${\bf q}$. 
The question of higher order corrections in 
the denominator expressing the cost in free energy related to the curvature
is still open \cite{Jacques, Napior}.\\
Curvature effects become much more dramatic when an amphiphilic film
is present at the interface because such a film may exhibit a large
bending stiffness and also decreases the surface tension. In microemulsions, 
for instance, surface tensions as low as $10^{-3} mN/m$ can be achieved, 
and curvature effects are dominant. The simplest phenomemological model 
due to Helfrich \cite{Helfrich} then predicts an additional  $K q^4$ 
term in the denominator of Eq. \ref{spectrum}. 
The bending rigidity $K$ will therefore become important in limiting the 
out-of-plane fluctuations only at scales smaller than $\sqrt{K/\gamma}$,
typically in the nanometer range for films spread at the air-water interface.
A comprehensive understanding of the bending rigidity in terms 
of molecular order and chain conformations is still lacking. This is 
a central problem in soft condensed matter physics where systems are often 
composed of films (monolayers or bilayers) and the role of 
fluctuations is dominant. 
The range of situations where the Helfrich Hamiltonian can be applied 
is also an open question from an experimental point-of-view. 
Indeed, the coupling of the out-of-plane fluctuations with the fluctuation
modes of any quantity of suitable symmetry may modify the energy spectrum
\cite{Peliti}.
For example, in the case of a tethered membrane having the elasticity 
of a solid, there can be a  coupling 
between out-of-plane fluctuations and in-plane modes due to the fact that 
many configurations of the film with low bending energy {\it per se} 
imply stretching. This leads to a divergence of $K(q)$ at large scale, or a rigidity term in the spectrum 
proportional to $ 1/{ q}^{4- \eta}, \> 1/2<\eta<1$ 
\cite{Nelson, Abraham,Petsche}.
Another interesting case is that of polarization 
for dipoles normal to the surface, always carried by classical amphiphiles,
which leads to a non-trivial $q^{-3}$ dependence\cite{Peliti}.\par

The examples cited in this introduction indicate that the key problems 
in liquid interfaces quite generally occur at very small length-scales. 
The many powerful techniques used to investigate solid surfaces
generally require ultra-high vacuum and cannot be applied to the study
of liquid surfaces. 
The scattering of short wave-length radiation (x-rays or neutrons) 
thus provides unique information.
The first reflectivity experiments on liquid surfaces were carried out
on liquid metals \cite{Srice, Bosio} in the 70's and the roughness of
the free surface of water was demonstrated to be consistent with the 
capillary wave model in the mid 80's \cite{Alan,11,old}. 
In reflectivity studies 
however, only density profiles averaged over the surface can be measured.
From the beginning of the 90's, the analysis of 
diffuse scattering from the surface, 
which gives access to the roughness spectrum has been 
undertaken\cite{Schwartz,Sanyal}. The spectrum could be measured 
up to wave-vectors of the order of $10^7 m^{-1}$. 
The method has been more recently extended down to molecular length-scales
giving access to new phenomena \cite{Chris}. After a short presentation 
of the method we discuss the results concerning the bare water surface 
and give some examples for surfactant monolayers.

\section{Experimental method.}

Langmuir films (monolayers of insoluble amphiphiles at the air/water 
interface) offer the possibility
of measuring the fluctuation spectra of 
oriented films with a well characterized structure. 
Temperature, molecular area or surface tension can indeed be fixed, 
the structure can be determined 
by grazing incidence x-ray diffraction and reflectivity, and the texture 
observed {\it in situ} by Brewster Angle Microscopy or by atomic
force microscopy (on transfered films). 
In this paper, we present direct measurements of  the fluctuation spectra  
by grazing-incidence diffuse x-ray scattering 
\cite{Sinha,nous,Kevin,Haase,Sun,Thomas}. \\
In a Langmuir film, 
the scattering of x-rays results from height fluctuations of  the different 
interfaces (air/chain, chain/head, head/subphase), assumed in this 
study to fluctuate conformally.  
The interferences between 
beams scattered at the different interfaces, which can be accounted for within 
the Distorted Wave Born Approximation taking into account 
refraction and absorption \cite{nous,Kevin}, give rise to the 
normal $q_z$ dependence. The $(q_x,q_y)$ dependence of the scattered
intensity is concentrated in the form:
\noindent $e^{-q_z^2<z^2>}\int dxdy[e^{q_z^2
<z(0,0)z(x,y)>}-1]e^{i(q_xx+q_yy)}\tilde R(x,y)$,
where $\tilde R(x,y)$ is the Fourier transform of the resolution function.
For small $q_z$, the exponential can be developped and one simply gets 
the convolution of the roughness spectrum
$<z({\bf q})  z({\bf -q})>$ with the resolution function.\\

The measurements were performed at the ``Tro\"{\i}ka'' and ``CRG-IF'' beamlines 
of the European Synchrotron Radiation Facility (ESRF).
Two types of experimental configurations were used (Fig.\ref{troika}). 
At the Troika beamline (undulator)
the desired radiation was selected using the (111)
reflection of a diamond monochromator in a Laue geometry. 
A SiC mirror was used in the monochromatic beam in order to fix the
grazing angle of incidence and eliminate higher order light. 
At the CRG-IF beamline (bending magnet) a double Si(111) monochromator
and a platinum coated glass mirror are used.
In both cases, the intensity of the $ 0.4 \times 0.2 mm^2$ ($w \times h$) beam
was approximately $2 \times 10^{10} counts/s$.
The two beamlines differ notably for these experiments
in the horizontal divergence of the
synchrotron x-ray beam at the sample.
A home-made Langmuir trough was mounted on 
a Nanofilm technologie GmbH active vibration isolation system
on the sample stage of the diffractometer.
Since the signal scattered by the interfacial fluctuations
at large in-plane wave-vectors $q_{xy}$ is very low ($\approx 10^{-10}$ 
of the incident beam), extreme care was taken to limit the background 
by shielding of the experimental device and by a flux of He gas in the Langmuir
trough. 
Another very important requirement is to fix the grazing angle of incidence
below the critical angle for total external reflection in order 
to limit the penetration and therefore scattering in the bulk.
Under total external reflection conditions the background could be 
measured by simply lowering the trough by $1mm$ 
and scanning around the direct beam.
The first type of experiments consisted of 
$\theta_d$ detector scans in the plane of incidence 
with a fixed angle of incidence $\theta_i$ (see Fig.\ref{troika}a).
With this geometry both $q_x = 2\pi/\lambda \left[cos(\theta_i)
- cos(\theta_d) \right]$ and $q_z = 2\pi/\lambda 
\left[sin(\theta_i) + sin(\theta_d) \right] \approx \sqrt{4 \pi q_x / \lambda}$
are varied jointly in an experiment.
In a second type of experiments (Fig.\ref{troika}b) 
we measured the scattered intensity in the $q_y$ direction, i.e. 
without any coupling to the vertical structure.
Intensity was collected using a NaI(Tl) scintillator in the first 
configuration and a gas-filled (xenon) 
position sensitive detector (PSD) in the second configuration.\\

\section{Bare water free surface.}

The intensity scattered by the bare water free surface
(from a Millipore purification system) is presented in (Fig.\ref{Wasser}) 
for both experimental configurations. In this case
the scattered intensity can be calculated without {\it any} adjustable 
parameter: the surface tension $\gamma_{H_2O}$ is known, as well as the 
different parameters determining the resolution of the experiment (slit
dimensions, incident and accepted beam footprints). 
The intensity scattered in the plane of 
incidence is presented in Fig.\ref{Wasser}a.
The data extend to $q \ge 10^8 m^{-1}$
and are well described by the spectrum of Eq.\ref{spectrum}. The data taken 
in the interface plane ($q_y$) extend to even larger wave-vector 
($ \approx 10^{10} m^{-1}$, note that this is more than two orders of 
magnitude than that of the best previous measurements \cite{Schwartz}) 
but can only be described with the spectrum 
of Eq.\ref{spectrum} up to $q \approx 5 \times 10^{8} m^{-1}$.
The excess scattering at $q \ge 10^{9} m^{-1}$ indicates either 
a smaller effective surface tension at such large wave-lengths \cite{Napior}
or another source of scattering. Because this scattering seems to have 
no dependence on $q$ (the apparent $q^{-1}$ dependence on Fig.\ref{Wasser} 
is due to the resolution function), it can be due to acoustic waves 
within the penetration depth of the beam. The corresponding scattering
cross-section \cite{Kevin} can be calculated from the density-density 
correlation function at or near the interface which can itself be determined 
by using the linear response theory \cite{Loudon}:
\begin{equation}
{d \sigma \over d \Omega} = {A (1-n^2)^2 \vert t_{in} \vert^2
\vert t_{out} \vert^2 k_B T  \kappa_T \over 2 {\cal I}m (q_z)} ,
\label{acoustic}
\end{equation}
where $n$ is the refraction index of water, $t_{in}$ and $t_{out}$ 
are the transmission coefficients of the air/water interface for the incident
and scattering beams, $\kappa_T$ is the isothermal compressibility 
of water ($4.58 \times 10^{-10} m^2 N^{-1}$), and 
${\cal I}m(q_z)$ is the imaginary part of the normal component of the 
wave-vector transfer  (inverse of the penetration length).
Including this contribution gives a better agreement (Fig.\ref{Wasser}b),
without discarding the possibility of other corrections.

\section{Surfactant monolayers.}
When a surfactant monolayer is present at the interface, its first effect 
is to reduce the surface tension: $\gamma = \gamma_{H_2O} 
- \Pi$, where $\Pi$ is the surface pressure,
as illustrated in Fig.\ref{scale} 
for an arachidic acid ($CH_3-(CH_2)_{18}-COOH$) film.
Higher order corrections to the spectrum, i.e.  effects of the bending
stiffness of the film are also evidenced.  
Results for a L$_\alpha$ di-palmitoylphosphatidylcholine
(DPPC) film on pure water are presented in Fig.\ref{DPPC}. 
Whereas at small $q_y$ values the scattered intensity scales with 
the surface tension as expected, this is no longer true at large $q_y$ due to
the effect of bending stiffness.  
We have analysed the data of Fig.\ref{DPPC} using the spectrum Eq.\ref{spectrum}
plus an additional term $K q^4$ in the denominator.
For the more compressed film of Fig.\ref{DPPC} we find $K = (5 \pm 2) k_BT$,
smaller than generally expected in condensed DPPC films \cite{Sackmann},
but consistent with a na\"{\i}ve estimation relying on the theory of the 
elasticity of solid shells: $K = \epsilon h^2 /12 \approx 
13 k_BT $ with $\epsilon \approx 0.3 N/m$ \cite{Albrecht}, 
the inverse of the compressibility
of the film of thickness $h \approx 1.4 nm$ \cite{Bayerl}.
The observed wave-vector range is not large enough to allow the precise 
determination on the exponent $4$.\\
We have observed exponents smaller than $4$  
in the case of the very rigid films\cite{Chris} formed by  
fatty acids (here behenic acid $CH_3-(CH_2)_{20}-COOH$)  
on divalent cation subphases 
($5 \times 10^{-4} mol/l$ $CdCl_2$) at high pH ($8.9$) 
and low temperature ($5^{\circ} C$). 
Uncompressed, such films exhibit a $q^{-3.3}$ power law 
which could be due to the coupling between in-plane (phonons) 
and out-of-plane elasticity \cite{Chris}.
Under compression, those rigid
films buckle in the third dimension as shown on Fig.\ref{CD}. The film 
is homogeneous at scales $< 1 \mu m$, and the only possible source of
scattering (i.e. refractive index inhomogeneities) are indeed surface
corrugations. As shown in Fig.\ref{CD}, the buckling wave-length 
$\Lambda \approx 2\pi / (3 \times 10^8 m^{-1}) \approx 20 nm$ changes
only slightly under compression, but the amplitude strongly increases.
This buckling could be induced by the asymmetry in packing constraints 
for headgroups and aliphatic chains, like in the ripple phase of
lyotropic liquid crystals \cite{Carlson} where similar periods are observed.
The buckling wave-length 
(Fig. \ref{CD}) could result from a balance between the headgroup size inducing 
spontaneous curvature, the chain length that would limit the possible 
curvature, surface tension and the energy cost of defects.\\

\section{Conclusion}

In conclusion, the possibility of measuring height fluctuation spectra 
of liquid-gas interfaces 
down to molecular scales, thus enabling the determination of exponents
and bending constants by means of x-ray surface scattering has been 
demonstrated. We are currently extending this method for measurements
of films at liquid-liquid interfaces.\\

$^*$ {\it present address:} LPMI, Universit\'e de Pau, 2 av. du pr\'esident 
Angot, F-64000 Pau.\\

\clearpage
\vfill
\eject
\newpage
\begin{figure}
\caption{Schematics of the experiment (Troika beamline)
(a): $(q_x,q_z)$ incidence plane geometry. (b): 
in-plane $q_y$ geometry. $C^*(111)$: diamond monochromator, SiC: mirror,
VP: vacuum paths, LT: Langmuir trough, 
NaI(Tl): scintillation detector. 
PSD: position sensitive gas-filled (xenon) detector.
The curve in (b) represents the scattered intensity (horizontal axis) 
as a function of the vertical position on the PSD.
%The sample-to-$S_4$ distance is 650 mm, and $S_3$-$S_4$ = 470mm.
%$S_1$ to $S_4$ are slits whose horizontal $\times$ 
%vertical openings were fixed to
%$S_1$: 0.6mm $\times$ 0.2mm,
%$S_2$: 0.4mm $\times$ 0.2mm,
%$S_3$: 2mm  $\times$ 1mm, 10mm or 22mm,
%$S_4$: 1mm  $\times$ 0.250mm.
}
\label{troika}
\end{figure}
\begin{figure}
\caption{Scattering by the bare water surface in the plane of
incidence (a) and in the plane of the surface (b). For measurements
in the plane of incidence $q_x$ is the projection of the wave-vector
transfer on the horizontal. ``$1 mm$'' (circles) 
and ``$10 mm$'' (triangles)
is the opening of slit $s_3$ in Fig.\ref{troika}a.
Squares represent the background which was not subtracted. In (b) 
the background was subtracted following the procedure indicated in text. 
The full line is calculated by using the capillary wave spectrum of 
Eq.\ref{spectrum} and the acoustic wave scattering cross section
Eq.\ref{acoustic} has been 
included to calculate the dashed line curve. Note that the great dynamic 
range of the y-scale has been compressed by a multiplication of the measured 
scattered intensity by $q_x$ or $q_y^2$.}
\label{Wasser}
\end{figure}
\vskip 1truecm
\begin{figure}
\caption{(a): Intensity scattered by an arachidic acid film (black curves) 
and water (grey curves). The surface tensions are (top to bottom) 
33 mN/m (diamonds), 43mN/m (triangles), 53mN/m (squares), 69mN/m (circles) and 73mN/m. (b): The same data normalized 
by $\gamma / \gamma_{water}$ in order to illustrate the scaling 
$I \propto \gamma$ in the range $3 \times 10^6 m^{-1} \le q_x \le 8 \times 
10^6 m^{-1}$ where capillary waves dominate the fluctuation spectra.
The fringes are due to the normal film structure since 
$q_z$ is not constant in the (x,z) configuration.}
\label{scale}
\end{figure}
\vskip 1truecm
\begin{figure}
\caption{Intensity scattered in the horizontal plane 
by a bare water surface (grey triangles) and a DPPC film at $3^{\circ} C$
compressed at $20 mN/m$ (grey circles) and $40 mN/m$ (black circles).
Lines are the best fits as indicated in text.
Note that the scattered intensity scales with the surface tension 
at low $q_y$ but that this is no longer true at large $q_y$ due to
the effect of bending stiffness (the black curve passes below the grey curves).
Inset: corresponding molecular area - surface pressure isotherm.}
\label{DPPC}
\end{figure}
\vglue 1truecm
\begin{figure}
\caption{Intensity scattered in the plane of incidence by an arachidic 
acid film deposited on a $5 \times 10^{-4} mol/l$ $Cdcl_2$ subphase
at $pH = 8.9$ and $T = 5^{\circ} C$. Bottom to top, $\Pi= 1.4 mN/m , \> 
1.8 mN/m , \> 5 mN/m , \> 14.6 mN/m , \> 17.3 mN/m$, and $  20 mN/m$. }
\label{CD}
\end{figure}
\vglue 18truecm
\clearpage
\vfill
\eject

\newpage

\epsfxsize 14truecm
\hfil \epsfbox{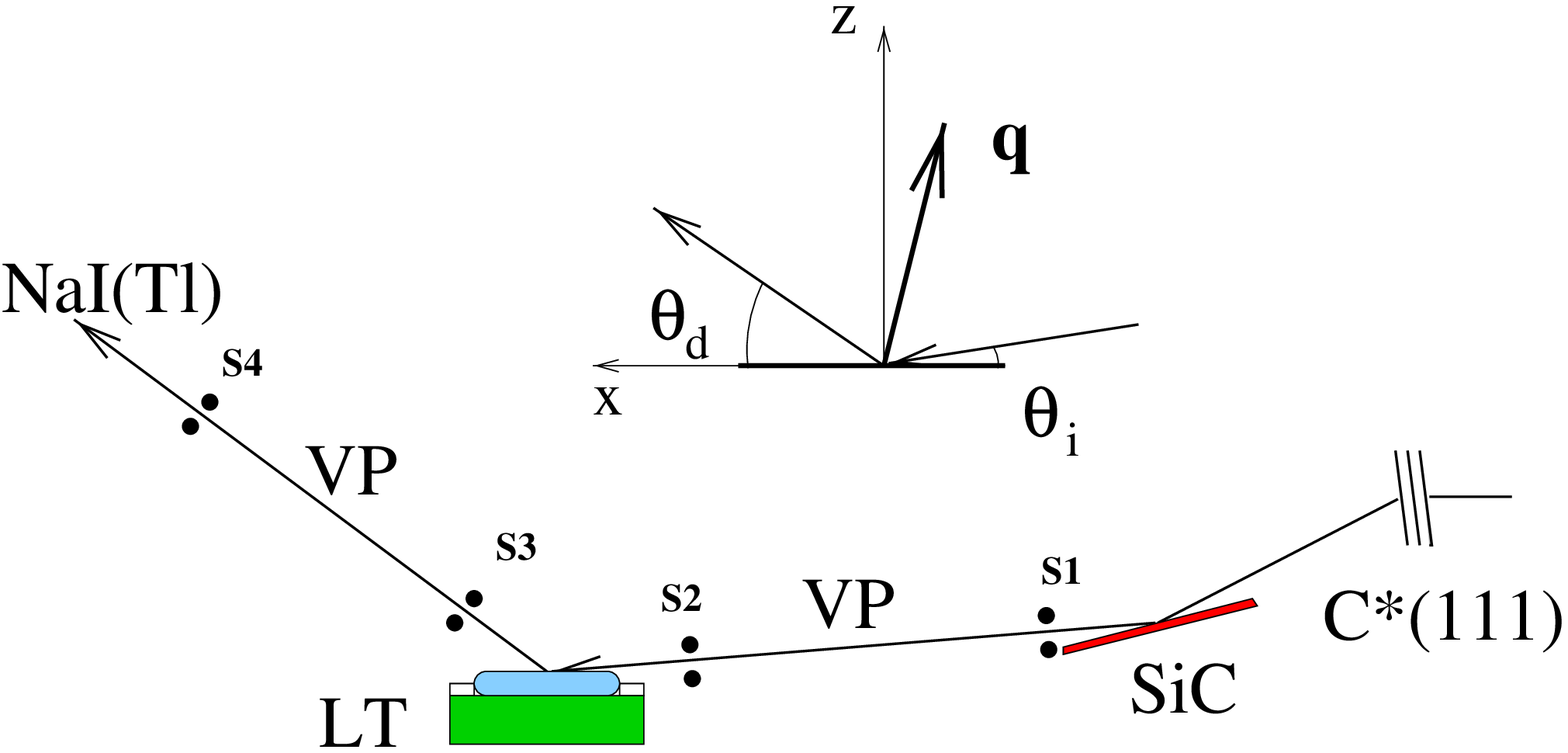} \hfil
\vglue .5truecm
\epsfxsize 14truecm
\hfil \epsfbox{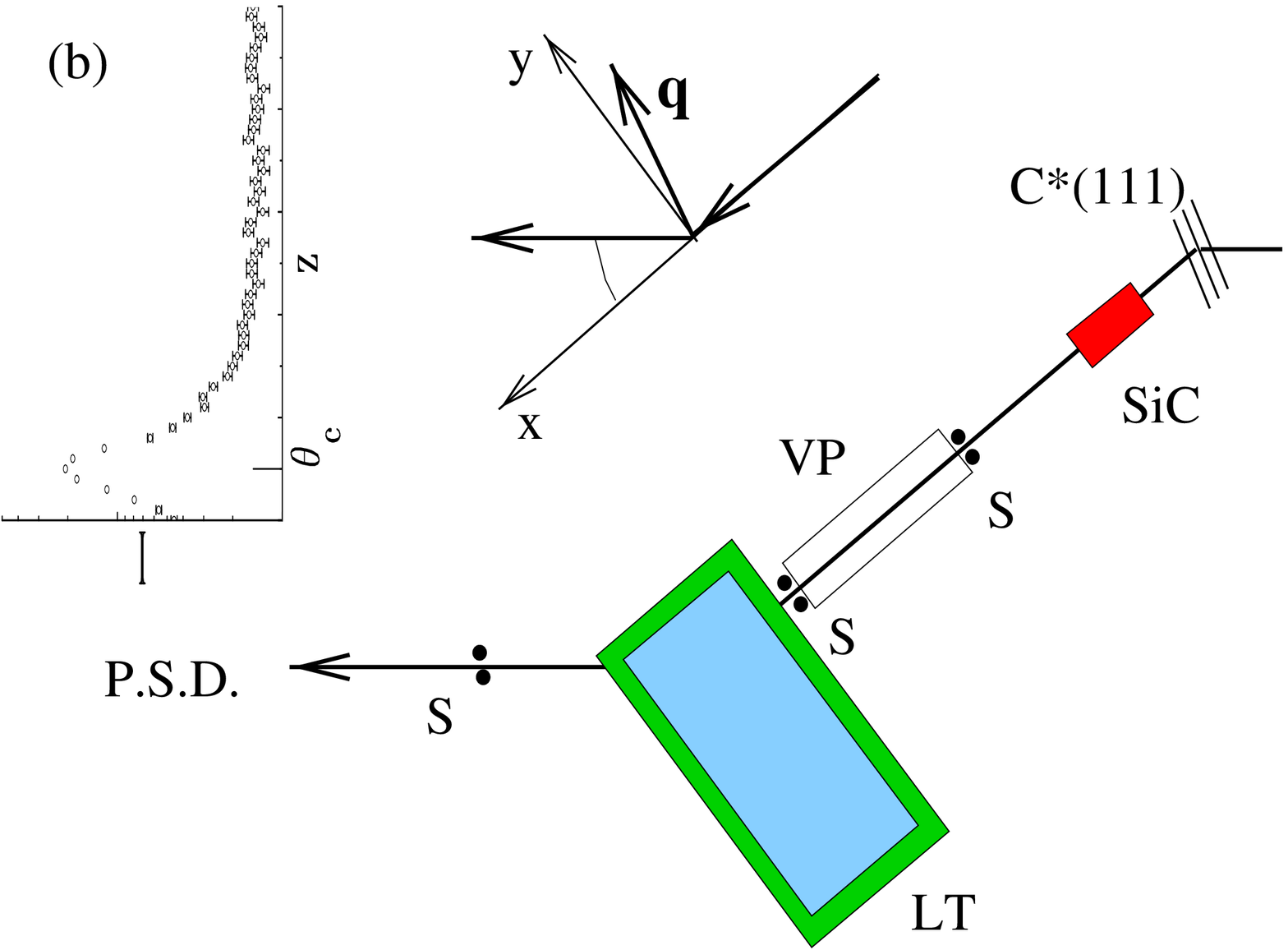} \hfil
\vskip 3truecm
\centerline{Fig. \ref{troika}}

\newpage
\epsfxsize 14truecm
\hfil \epsfbox{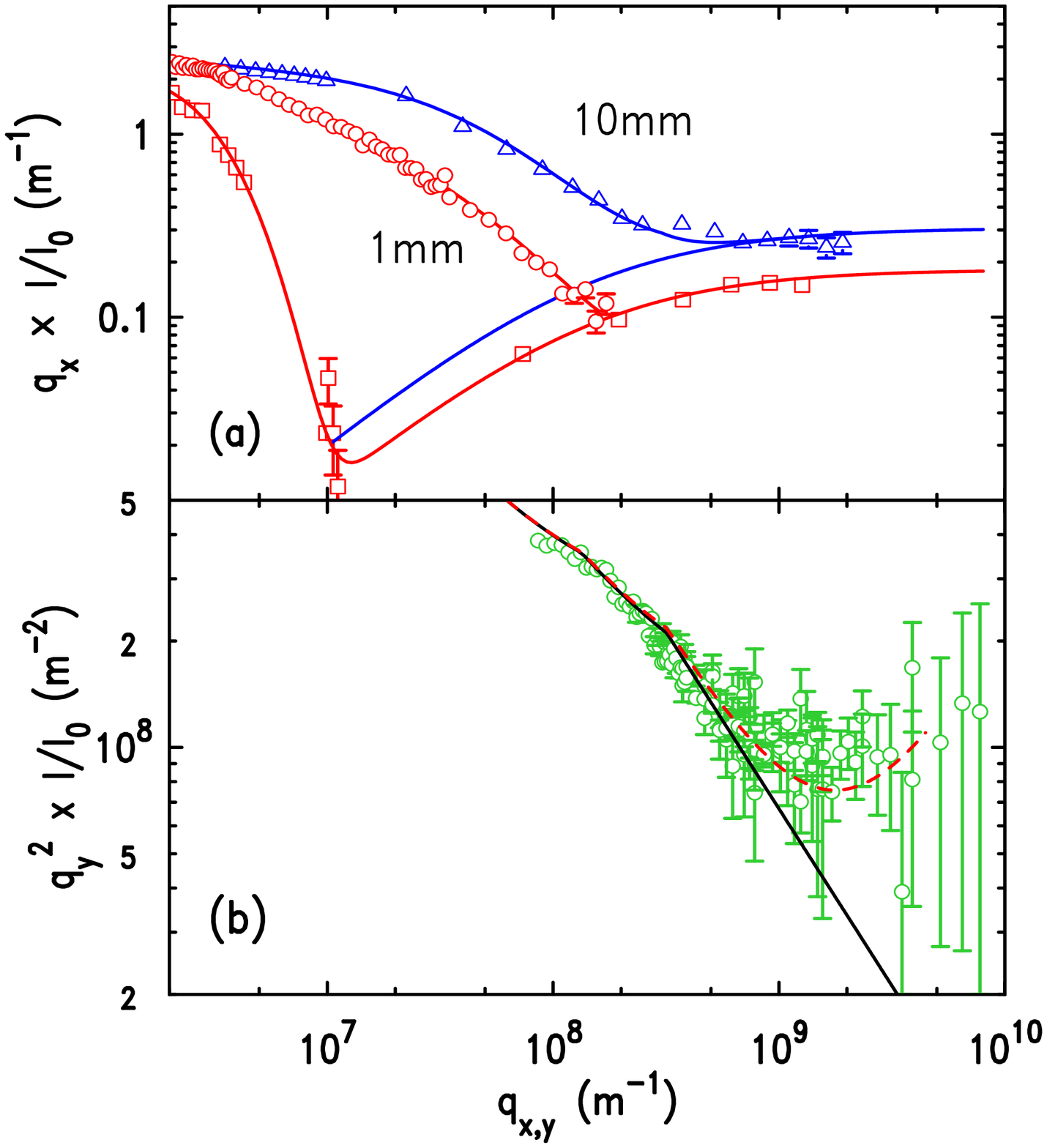} \hfil
\vskip 3truecm
\centerline{Fig. \ref{Wasser}}

\newpage
\epsfxsize=14truecm
\hfil \epsfbox{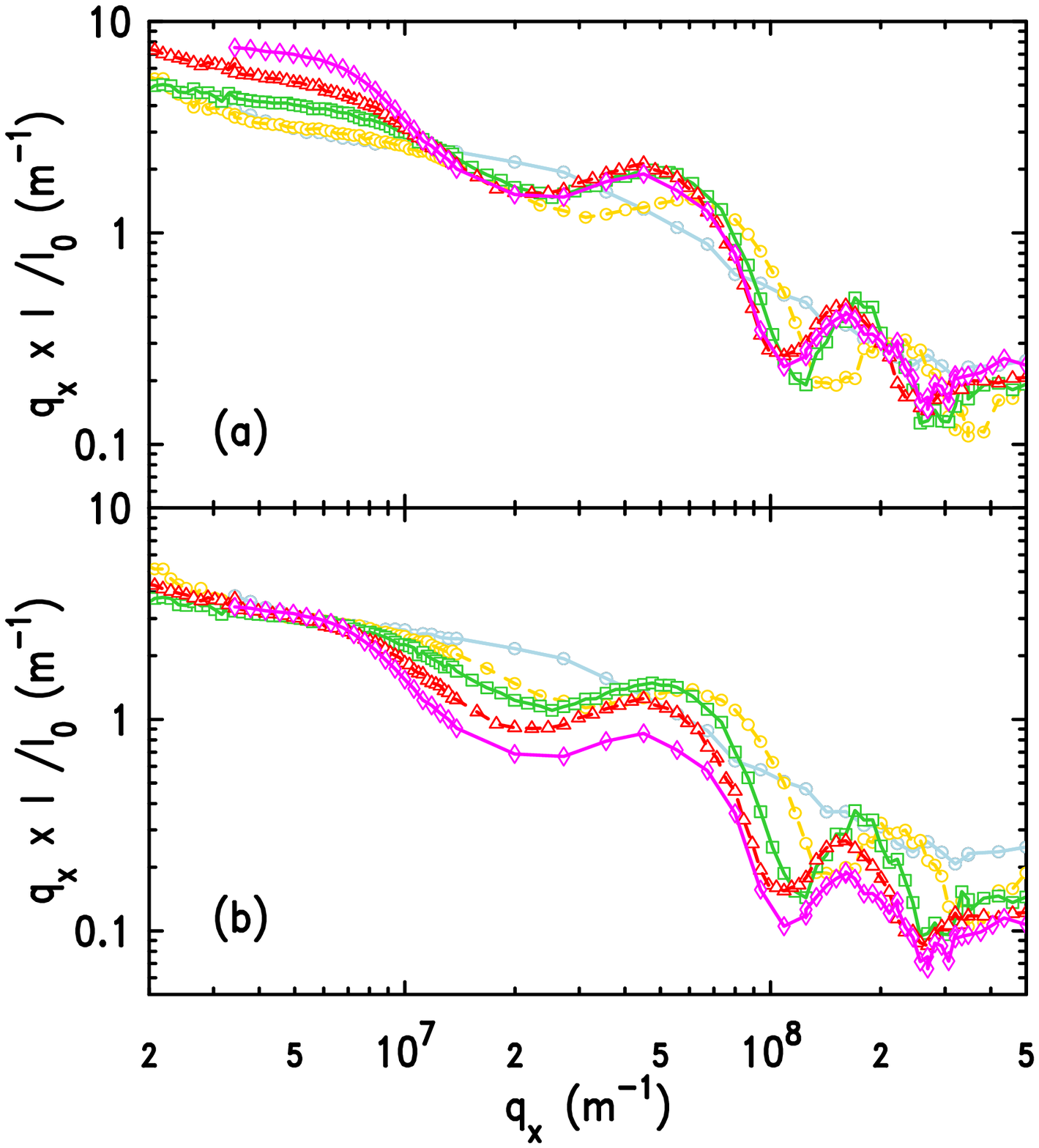} \hfil
\vskip 3truecm
\centerline{Fig. \ref{scale}}

\newpage
\epsfxsize=14truecm
\hfil \epsfbox{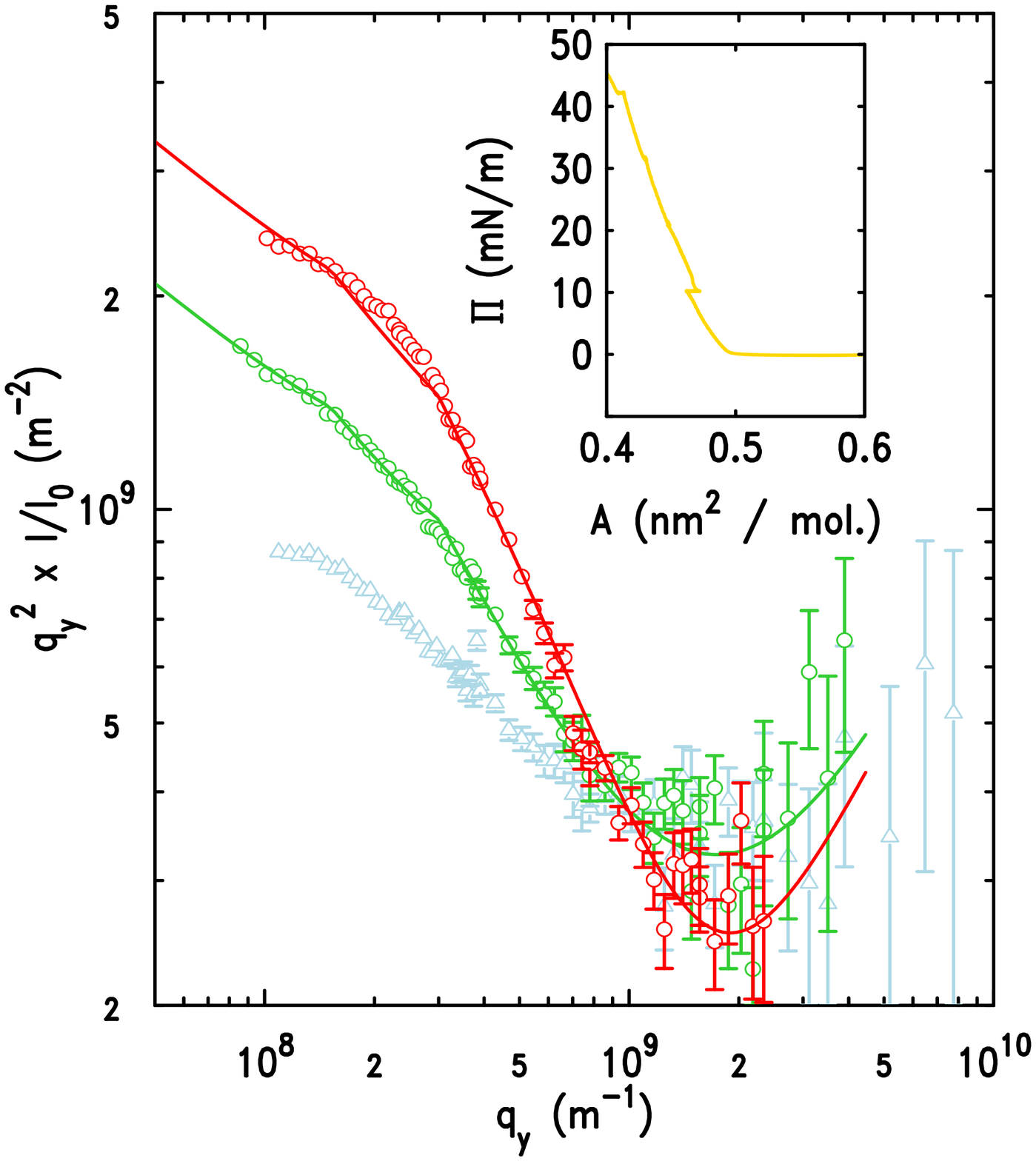} \hfil
\vskip 3truecm
\centerline{Fig. \ref{DPPC}}

\newpage
\epsfxsize=14truecm
\hfil \epsfbox{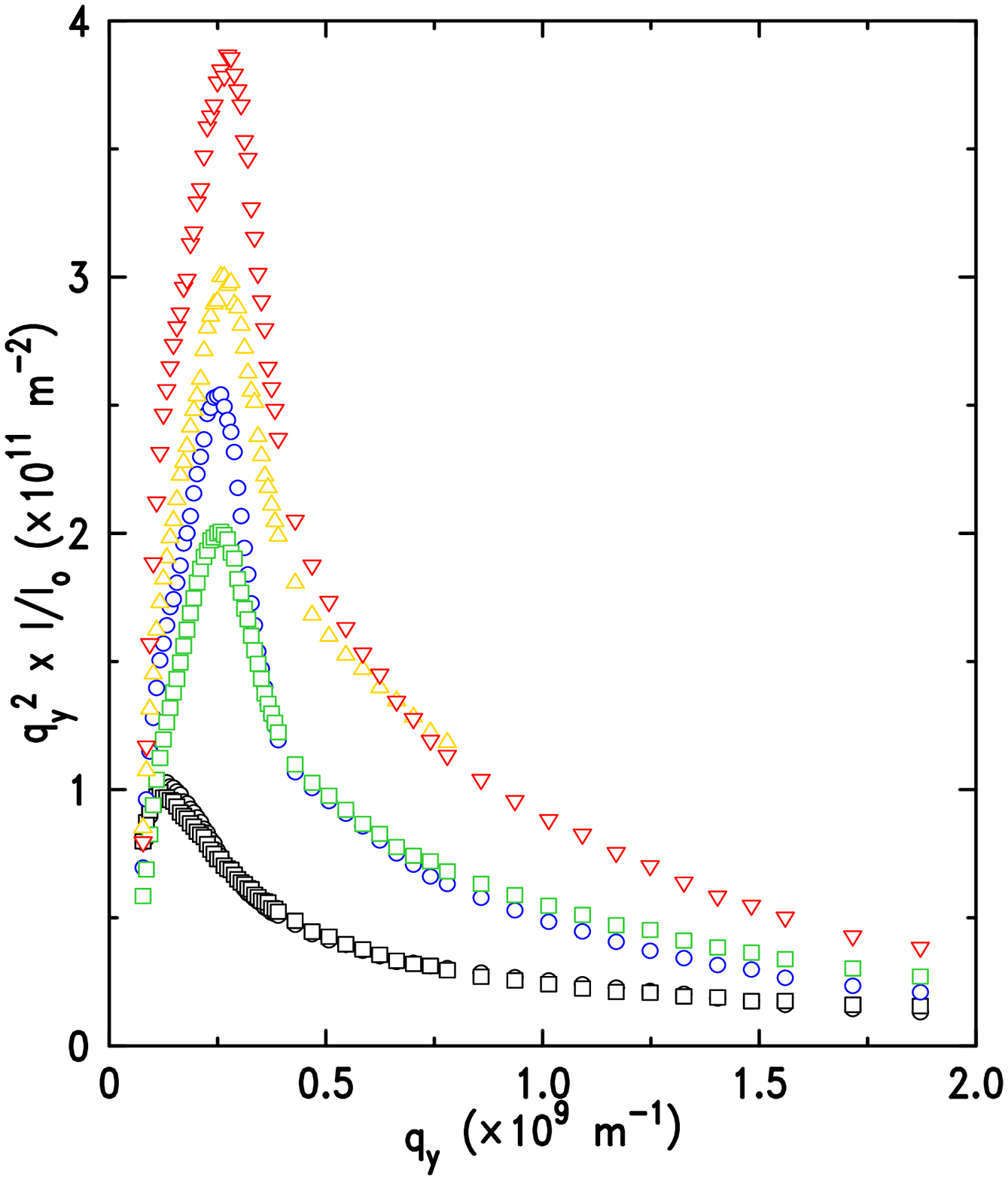} \hfil
\epsfxsize=8truecm
\vglue -14truecm
\hglue 6truecm \epsfbox{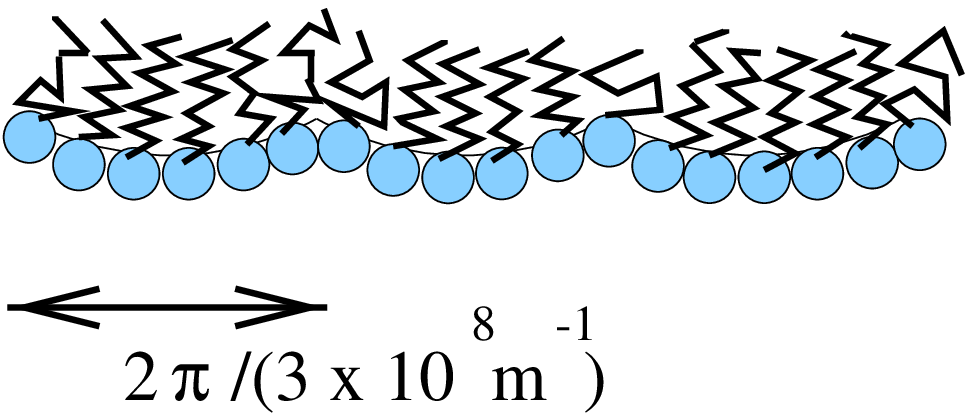}
\vglue 10truecm
\vskip 3truecm
\centerline{Fig. \ref{CD}}


\newpage

\begin{thebibliography}{9}
\parsep -.3truecm
\itemsep -.2truecm
\bibitem{Widom}  J.S. Rowlinson and B. Widom, ``Molecular 
Theory of Capillarity''  Clarendon Press, Oxford 1982.
\bibitem{Buff} F.P. Buff, R.A. Lovett and F.H. Stillinger Jr. {\it 
Phys. Rev. Lett.} {\bf 15} 621 (1965).
\bibitem{Loudon} See for example R. Loudon, ``Ripples on liquid interfaces''
in ``Surface excitations'' edited by V.M. Agranovich and R. Loudon,
Modern Problems in Condensed Matter Science, vol. 9, North-Holland, 
Amsterdam, 1984.
\bibitem{Jacques} J. Meunier, {\it J. Physique} {\bf 48} 1819 (1987).
\bibitem{Napior} M. Niap\'orkowski, S. Dietrich, {\it Phys. Rev. E} 
{\bf 47} 1836 (1993).
\bibitem{Helfrich} W. Helfrich, {\it Z. Naturforschung} {\bf 28 c} 693 (1973).
\bibitem{Peliti} L. Peliti and J. Prost, {\it J. Phys. France} {\bf 50} (1989) 1557.
%\bibitem{Jacques0} J. Meunier {\it J. Phys. France} {\bf 48} (1987) 1819. 
%\bibitem{Jacques} L. Peliti and S. Leibler, {\it Phys. Rev. Lett.} {\bf 54} (1985) 1690.
\bibitem{Nelson} D.R. Nelson and L. Peliti, {\it J. Phys. France} {\bf 48} (1987) 1085. 
\bibitem{Abraham} F.F. Abraham and D.R. Nelson {\it Science} {\bf 249} (1990) 393. 
\bibitem{Petsche} R. Lipowsky and M. Girardet, {\it Phys. Rev. Lett.} {\bf 65} (1990) 2893;
F.F. Abraham {\it Phys. Rev. Lett.} {\bf 67} (1991) 1669;
P. Le Doussal and L. Radzihovsky {\it Phys. Rev. Lett.} {\bf 69} (1992) 1209;
I.B. Petsche and G.S. Grest {J. Phys. I France} {\bf 1} (1993) 1741.
\bibitem{Srice} B.C. Lu, S.A. Rice, {\it J. Chem. Phys.} {\bf 68} 5558 (1978).
\bibitem{Bosio} L. Bosio, M. Oumezine, {\it J. Chem. Phys.} {\bf 80} 
959 (1984).
\bibitem{Alan} A. Braslau, M. Deutsch, P.S. Pershan, A.H. 
Weiss, J. Als-Nielsen, J. Bohr, {\it Phys. Rev. Lett.} {\bf 54} 114 (1985).
\bibitem {11}A. Braslau, P.S. Pershan, G. Swislow, B.M. Ocko and J. 
Als-Nielsen, {\sl Phys. Rev. A\/}, {\bf 38}, 2457 (1988).
\bibitem{old} J. Daillant, L. Bosio, B. Harzallah, J.J. Benattar, {\it J. Phys. France II} 
{\bf 1} 149 (1991). 
\bibitem{Schwartz} D.K. Schwartz, M.L. Schlossman, E.H. Kawamoto, G.J. Kellog, and P.S. Pershan, 
{\sl Phys. Rev. A} {\bf 41} 5687 (1990). 
\bibitem{Sanyal} M.K. Sanyal, S.K. Sinha, K.G. Huang, B.M. Ocko, 
{\it Phys. Rev. Lett.} {\bf 66} 628 (1991).
\bibitem{Chris} C. Gourier, J. Daillant, A. Braslau, M. Alba, K. Quinn, 
D. luzet, C. Blot, D. Chatenay, G. Gr\"ubel, J.-F. Legrand, and G. Vignaud,
{\it Phys. Rev. Lett.} {\bf 78} 3157 (1997).
\bibitem{Sinha} S.K. Sinha, E.B. Sirota, S. Garoff, and H.B. Stanley,  
Phys. Rev. B {\bf 38}, 2297 (1988).
\bibitem{nous} J. Daillant, O. B\'elorgey {\it J. Chem. Phys.} {\bf 97}
5824 (1992).
\bibitem{Kevin} J. Daillant, K. Quinn, C. Gourier, F. Rieutord, 
{\it J. Chem. Soc. Faraday Trans.}, {\bf 92} 505.
\bibitem{Haase} S. Dietrich, A. Haase, {\it Physics Reports}, {\bf 260}
1 (1995).
\bibitem{Sun} S.K. Sinha {\it Current opinion in solid state and
material science} {\bf 1} 645 (1996).
\bibitem{Thomas} R.K. Thomas, J. Penfold {\it Current opinion in colloid and 
interface science} {\bf 1 } 23 (1996).
\bibitem{Sackmann} E. Sackmann in ``Handbook of biological physics'', vol. 1A
edited by R. Lipowsky and E. Sackmann, Noth-Holland, Amsterdam, 1995.
\bibitem{Albrecht} O. Albrecht, H. Gruler, and E. Sackmann, {\it J. Phys.
France} {\bf 39} 301 (1978).
\bibitem{Bayerl} T.M. Bayerl, R.K. Thomas, J. Penfold, A. Rennie, and 
E. Sackmann, {\it Biophys. J.} {\bf 57} 1095 (1990).
{\it Phys. Rev. Lett.} {\bf 78} 3157 (1997).
%\bibitem{Laurent} L. Bourdieu, J. Daillant, D. Chatenay, A. Braslau, and 
%D. Colson, {\sl Phys. Rev. Lett.} {\bf 72}, 1502 (1994).
\bibitem{Carlson} J.M. Carlson and J.P. Sethna, {\it Phys. Rev. A} {\bf 36}
3359 (1987).

\end{thebibliography}
\end{document}